\documentclass[runningheads]{llncs}
\usepackage[utf8]{inputenc}
\usepackage{hyperref}
\usepackage{algorithm}
\usepackage{algpseudocode}
\usepackage{amsmath}  
\usepackage{amssymb} 
\usepackage{booktabs}
\usepackage{colortbl}
\usepackage{color}
\usepackage{comment}
\usepackage{eurosym}
\usepackage{glossaries}
\usepackage{graphicx}
\usepackage{lastpage}
\usepackage{listings}
\usepackage{longtable}
\usepackage{multicol}
\usepackage{multirow}
\usepackage{soul}
\usepackage{subfigure}
\usepackage{textcomp}
\usepackage{url}
\usepackage{xcolor}
\usepackage{tikz}
\usetikzlibrary{arrows.meta, positioning, shapes.geometric, backgrounds, fit, calc}
\usepackage{svg}

\title{vSPACE: Voting in a Scalable, Privacy-Aware and Confidential Election\thanks{Supported by Innovate UK.}}

\author{Se Elnour\inst{1} \and
William J Buchanan\inst{1} \and Paul Keating\inst{2} \and Mwrwan Abubakar\inst{1} \and Sirag Elnour\inst{3} }
\authorrunning{Elnour et al.}

\institute{%
    Blockpass ID Lab, Edinburgh Napier University, Edinburgh, UK\\
    \email{\{s.elnour,b.buchanan\}@napier.ac.uk} \and
    Democracy Counts\\
    \email{paulkeating@democracycounts.co.uk} \and
    vSpace Wallet\\
    \email{vote@vspace.wallet}
}

\begin{document}
%
%
{\def\addcontentsline#1#2#3{}\maketitle}
%

\vspace{-0.5cm}
\begin{abstract}
The vSPACE experimental proof-of-concept (PoC) on the [True{\bf\textit{Elect}}][{\bf\textit{Anon}}Creds] protocol presents a novel approach to secure, private, and scalable elections, extending the TrueElect and ElectAnon protocols with the integration of AnonCreds SSI (Self-Sovereign Identity). Such a protocol PoC is situated within a Zero-Trust Architecture (ZTA) and leverages confidential computing, continuous authentication, multi-party computation (MPC), and well-architected framework (WAF) principles to address the challenges of cybersecurity, privacy, and trust over IP (ToIP) protection. Employing a Kubernetes confidential cluster within an Enterprise-Scale Landing Zone (ESLZ), vSPACE integrates Distributed Ledger Technology (DLT) for immutable and certifiable audit trails. The Infrastructure as Code (IaC) model ensures rapid deployment, consistent management, and adherence to security standards, making vSPACE a future-proof solution for digital voting systems. 

\end{abstract}
\vspace{-0.25cm}

\vspace{-0.5cm}
\section{Introduction}
\vspace{-0.25cm}
Electronic voting is frequently employed in various decision-making elections since it is adaptable, easy to use, and inexpensive compared to a traditional poll \cite{singh2022review}. Despite this, the currently used electronic voting techniques pose the risk of excessive power and the manipulation of information, which reduces the basic levels of fairness, privacy, secrecy, anonymity, and transparency in the voting process. However, there are a variety of approaches that have been used in electronic and online voting systems. These methods use various encryption and decryption strategies to ensure that data exchanges are kept safe.

\begin{figure*}
\centering
\includegraphics[width=1.0\textwidth]{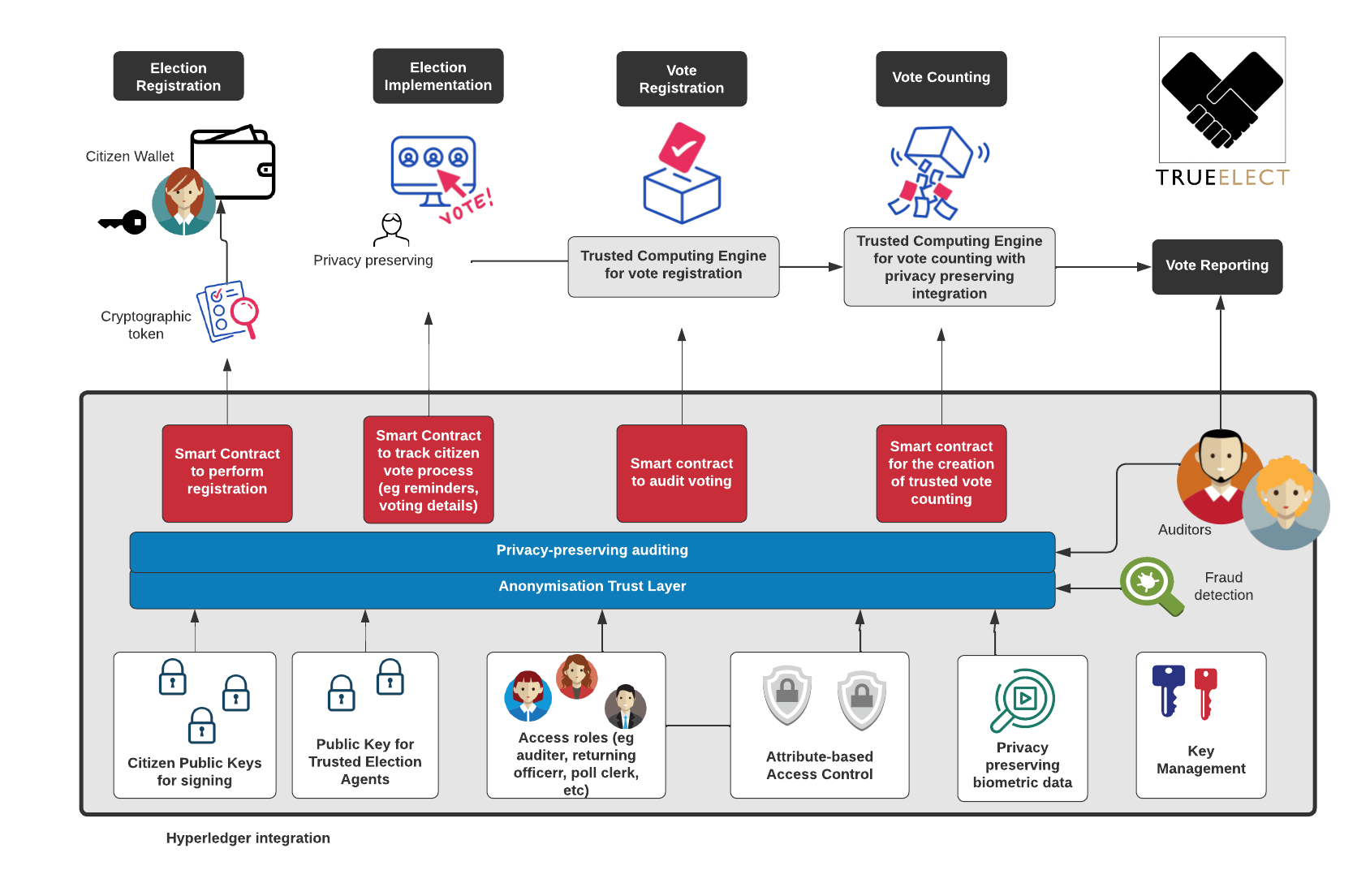}
\vspace{-0.25cm}
\caption{TrueElect Voting Outline, credit: asecuritysite.com/blogs/graphics}
\vspace{-0.25cm}
\label{fig:TrueElectAnonCreds__TrueElect_Protocol_Flow_Timeline}
\end{figure*}

The [True{\bf\textit{Elect}}][{\bf\textit{Anon}}Creds] vSPACE (Voting in a Scalable, Privacy-Aware and Confidential Election) is a novel solution that extends our TrueElect [as shown in the below Figure \ref{fig:TrueElectAnonCreds__TrueElect_Protocol_Flow_Timeline}] and ElectAnon \cite{onur2022electanon} Protocols, leveraging confidential computing, continuous authentication, multi-party computation, and well-architected framework (WAF) principles. This paper presents an experimental proof-of-concept on Self-Sovereign Identity (SSI), using Zero-Trust Architecture (ZTA) of Critical Infrastructure as Code (IaC) for cybersecurity, privacy, and Trust Over IP (ToIP) protection.

The main contribution of this paper is the design and implementation of a Kubernetes confidential cluster within an Enterprise-Scale Landing Zone (ESLZ) and which integrates Distributed Ledger Technology (DLT) for immutable and certifiable audit trails. This approach addresses the challenges of existing voting systems - offering a high level of security, privacy, and trust for both voters and election authorities.

\vspace{-0.5cm}
\vspace{-0.25cm}

\vspace{-.25cm}
\section{Related work}
\label{sec:secure-DLT-voting-requirements}
\vspace{-0.25cm}
The development of secure Distributed Ledger Technology (DLT)-based voting systems is contingent upon meeting a comprehensive set of requirements. These criteria are essential for preserving the integrity, confidentiality, and resilience of the voting process, thereby ensuring voter trust and the legitimacy of election outcomes.

Notwithstanding that, to analyze \textit{Eligibility, Uniqueness, Privacy, Universal Anonymity, Fairness, Accuracy, Universal Verifiability, Individual Verifiability, Robustness, Autonomy,} and \textit{Scalability}, Onur \& Yurdakul (2022) \cite{onur2022electanon} evaluated four prominent DLT-voting protocols [McCorry et al. (2017) \cite{ovnet}, Chaintegrity (2019) \cite{Zhang2019ChaintegrityBL}, Yang et al. (2020) \cite{YANG2020859}, and Panja et al. (2020) \cite{panja}]; nonetheless, this section analyzes such necessary requirements for a future-proof DLT-based voting as analyzed by Onur \& Yurdakul (2022) \cite{onur2022electanon}, and given that the security of online elections has been studied as early as 2002 by D.A Gritzalis \cite{GRITZALIS2002539}, and in 2021, Jafar et al work \cite{sensors} mentioned requirements for secure DLT-based online elections; such as: Eligibility, Uniqueness, Privacy, Universal Anonymity, Fairness, and Accuracy.

Moreover, these foundational requirements guide the design and implementation of our secure DLT-voting system, ensuring the integrity and trustworthiness of the election process; and such core requirements for secure DLT-voting are outlined as::
\begin{itemize}
\item \textbf{Eligibility}: Ensuring that only authorized individuals participate in the election is paramount. The system must incorporate mechanisms for verifying voter eligibility without compromising their privacy \cite{GRITZALIS2002539}.
Only eligible voters should vote \cite{GRITZALIS2002539}. {\bf \textit{Eligibility}} requires trusting the authority. DLT-voting is vulnerable if not properly implemented, enabling vote-buying and coercion.
\item \textbf{Uniqueness}: To uphold the principle of "one person, one vote," the system must prevent double voting by ensuring each eligible voter can cast only one vote \cite{GRITZALIS2002539}.
Each voter should cast one ballot \cite{GRITZALIS2002539} \cite{YANG2020859}. {\bf \textit{Uniqueness}} eliminates double voting.
\item \textbf{Privacy}: The confidentiality of a voter's ballot is crucial. The system must safeguard the secrecy of the vote, preventing any party from determining a voter's choices \cite{GRITZALIS2002539}.
Privacy may enable multiple votes from one voter using different keys \cite{ovnet} \cite{panja}. Some protocols bind keys to eligibility certificates to prevent this. No results should be available before tallying \cite{GRITZALIS2002539} \cite{sensors}. 
\item \textbf{Fairness}: The voting process should be conducted in a manner that prevents the premature disclosure of results, thereby ensuring a fair election. The system must also secure votes against any post-casting alterations \cite{GRITZALIS2002539}.
{\bf \textit{Fairness}} avoids biased voting. Protocols relying on authorities' or candidates' keys risk attacks to learn intermediate results \cite{Zhang2019ChaintegrityBL}. Candidates may not cooperate if unsatisfied \cite{YANG2020859}.
\item \textbf{Accuracy}: It is critical that every vote is accurately recorded and counted in the final tally, with the system being resilient against vote manipulation or fraud \cite{GRITZALIS2002539}.
{\bf \textit{Accuracy}} discards invalid votes. Usually zero-knowledge proofs ensure validity. Some protocols make invalid votes impossible \cite{ovnet} \cite{panja}. Only valid ballots should count \cite{accuracy}.
\item \textbf{Individual Verifiability}: Voters should have the capability to confirm that their vote has been correctly recorded and included in the final tally, enhancing their confidence in the voting process \cite{GRITZALIS2002539}.
\item \textbf{Universal Verifiability}: The integrity of the election process must be verifiable by any observer, ensuring that the system provides means for third parties to independently verify the correct counting of votes \cite{GRITZALIS2002539}.
\item \textbf{Universal Anonymity}: In addition to privacy, the system must ensure that a voter's identity cannot be linked to their voting behavior or other DLT activities, thus guaranteeing universal anonymity \cite{GRITZALIS2002539}.
Voters' addresses \cite{ovnet} \cite{panja} or public keys \cite{Zhang2019ChaintegrityBL}, \cite{YANG2020859}, \cite{priscore} are used. Both approaches risk \textit{linkage attacks}. 
\item \textbf{Robustness}: The voting system must demonstrate resilience against a broad spectrum of potential attacks and failures, ensuring the election's smooth progression under various conditions \cite{GRITZALIS2002539}.
\item \textbf{Scalability}: The system must efficiently manage a large number of voters and candidates without compromising performance or security, enabling scalability according to election needs \cite{GRITZALIS2002539}.
\end{itemize}

\vspace{-0.1cm}
\section{Design}
\vspace{-0.25cm}
\label{sec:the-TrueElectanoncreds-protocol-design}
The [True{\bf\textit{Elect}}][{\bf\textit{Anon}}Creds] protocol is a novel approach designed to address the challenges of scalability, privacy, and confidentiality in distributed ledger technology (DLT)-based voting systems. Its design incorporates the principles of privacy-preserving auditing and anonymization trust layers, as depicted in the provided diagram. It utilizes smart contracts for various stages of the election process, including voter registration, vote auditing, and the creation of trusted vote counting. The integration of Hyperledger ensures a secure and transparent process, with public keys for citizens and trusted election agents, as well as attribute-based access control for various roles such as auditors and returning officers. Privacy-preserving biometric data and key management are also critical components of the protocol, ensuring the security and confidentiality of the voting process.

Such [True{\bf\textit{Elect}}][{\bf\textit{Anon}}Creds] protocol proposal extends the capabilities of our own {\bf\textit{TrueElect}} protocol, as well as the {\bf\textit{ElectAnon}} protocol, as proposed by Onur \& Yurdakul (2022); with the integration of Hyperledger {\bf\textit{AnonCreds}} for Self-Sovereign Identity (SSI) to be implemented by our vSpaceWallet; alongside our vSpaceVote Confidential Computing cluster, leveraging ZKPs (Zero-Knowledge Proofs to ensure voter anonymity while maintaining the integrity and verifiability of the election-tech process.

Noting that the TrueElect and ElectAnon protocols were designed to facilitate secure and private voting mechanisms. Both protocols employ advanced cryptographic techniques to ensure the integrity and confidentiality of the voting process. However, they differ in their specific implementations and the flow of their respective protocols.
\vspace{-0.25cm}
\subsection{Overview of EMS, SSI, ZTA, IaC, and ToIP}
\vspace{-0.25cm}
The Election Management System (EMS) integrated with Self-Sovereign Identity (SSI), Zero-Trust Architecture (ZTA), Infrastructure as Code (IaC), and Trust Over IP (ToIP) represents a paradigm shift in the domain of digital elections. This integration aims to establish a certifiable, privacy-aware, and scalable framework for conducting elections, leveraging the latest advancements in cybersecurity and digital identity management.

\begin{itemize}
\item \textbf{Election Management System (EMS)}: The EMS is a comprehensive system designed to manage all aspects of the electoral process, from voter registration to ballot casting and result tabulation. The goal is to ensure a secure, transparent, and efficient election process that can scale to accommodate a large number of constituents while maintaining the integrity of the vote.
\item \textbf{Self-Sovereign Identity (SSI)}: SSI is a user-centric approach to digital identity that gives individuals control over their personal data. In the context of EMS, SSI allows constituents to manage their digital identities securely and to interact with the election system without relinquishing personal data control to third parties.
\item \textbf{Zero-Trust Architecture (ZTA)}: ZTA is a security model that operates on the principle of "never trust, always verify." It is designed to protect against internal and external threats by rigorously verifying the identity and trustworthiness of every device and user before granting access to the system's resources.
\item \textbf{Infrastructure as Code (IaC)}: IaC is a method of managing and provisioning computing infrastructure through machine-readable definition files, rather than physical hardware configuration or interactive configuration tools. This approach enables the rapid and consistent deployment of infrastructure, which is essential for the scalability and reliability of the EMS.
\item \textbf{Trust Over IP (ToIP)}: ToIP is a set of principles and standards that aim to establish a global framework for digital trust. It combines governance, technology, and legal considerations to create a secure, decentralized system for digital interactions. Within the EMS, ToIP ensures that all communications and transactions are secure and that the privacy of the constituents is preserved.
\end{itemize}


\vspace{-0.25cm}
\subsection{Purpose and Scope of the Proof of Concept}
\vspace{-0.25cm}
The Proof of Concept (PoC) is designed to validate the practical application and effectiveness of integrating an Election Management System (EMS) with Self-Sovereign Identity (SSI), Zero-Trust Architecture (ZTA), Infrastructure as Code (IaC), and Trust Over IP (ToIP) for the purpose of enhancing the security, privacy, and scalability of digital elections. The PoC aims to demonstrate the feasibility of this integrated approach and to provide a blueprint for its implementation in real-world election scenarios.

The scope of the PoC encompasses several key components:
\begin{itemize}
    \item The creation of a digital ballot within a secure and verifiable EMS, as outlined in the Business Process Model and Notation (BPMN) diagrams provided.
    \item The execution of a verification algorithm that ensures the integrity of the digital ballot and the eligibility of the voters, as depicted in the BPMN and sequence diagrams.
    \item The integration of a constituent digital wallet, which allows voters to securely manage their digital identities and interact with the election system using SSI principles within a ZTA framework.
    \item The application of IaC to facilitate the rapid and consistent deployment of the necessary infrastructure to support the EMS, while adhering to ToIP standards for secure and private digital interactions.
\end{itemize}

The PoC serves to illustrate the technical viability and the potential benefits of our adopted framework, such as improved election security, enhanced voter privacy, and the ability to scale to accommodate a large electorate. It also seeks to address the challenges associated with digital elections, including the prevention of fraud, ensuring the verifiability of votes, and maintaining the confidentiality of voters' choices.


\vspace{-0.1cm}
\section{Implementation}
\vspace{-0.25cm}

\begin{algorithm}
\caption{\textit{MPC (Multi-Party Computation)} integration between business processes and distributed registry \textit{DLT (Distributed Ledger Technology)}}
\begin{algorithmic}[1]

\State Constituency Acting Returning Officers or \textbf{\textit{POs (Presiding Officers)}}:
\State Initiate consortium's distributed \textit{EMS (Elections Management System)}
\State Finalise security tokens' \textit{HYOK (Hold Your Own Key)} ceremony
\State Activate confidential \textit{ZTA (Zero-Trust Architecture)} integrity attestation
\State Verify \textit{DID (decentralized identity)} with \textit{SSI (Self-Sovereign Identity)} Wallet
\State Enable \textit{MFCA (Multimodal Fusion-based Continuous Authentication)}
\State Multi-sign constitution counterparts irrevocably as Ricardian contracts
\State Proceed with overseeing an anonymised \textit{ElectionProcess} if authenticated

\For{each \textit{ElectionProcess} in \textit{EMS}}
    \State Generate a smart contract for the election process
    \If{\textit{ElectionContractAddress} is set}
        \State Generate a variable containing hard coded address
    \Else
        \State Generate a process contract constructor parameter for setting the election contract address
    \EndIf
\EndFor

\For{each \textit{VoterRegistration} in \textit{EMS}}
    \State Obtain voter \textit{DID}, eligibility proof from \textit{SSI-as-a-Service}
    \State Generate Chaincode or Solidity code for registering voter on the \textit{DLT}
    \State Ensure verification of identity using AI-based biometric \textit{MFCA}
    \State Inject generated \textit{DID} code to function body of registration process
\EndFor

\For{each \textit{VoteCasting} in \textit{EMS}}
    \State Obtain voter identity, ballot choice from \textit{SSI-as-a-Service}
    \State Generate Chaincode or Solidity code for casting vote on the \textit{DLT}
    \State Ensure privacy-preserving verification of voter identity and eligibility
    \State Inject generated voting code to function body of voting process
\EndFor

\For{each \textit{VoteTallying} in \textit{EMS}}
    \State Aggregate votes using secure \textit{MPC} and the quantum-safe \textit{DLT}
    \State Generate Chaincode or Solidity code for tallying votes
    \State Ensure \textit{ZKP (Zero-Knowledge Proof)} verification of integrity attestation
    \State Inject generated tallying code to function body of tallying process
\EndFor

\For{each \textit{VoteCertifiability} in \textit{EMS}}
    \State Verify the certifiable audited integrity and authenticity of the votes
    \State Use registry services \textit{ZTA} to validate vote transactions
    \State Update the \textit{EMS} with the verified vote counts to be announced
    \State Multi-sign confidential consensus by a quorum of \textit{POs} if verified
    \State Publish combined certifiable \textit{ZKPs} and audit into public \textit{DLTs}
\EndFor

\end{algorithmic}
\end{algorithm}

The executed verification algorithm is a pivotal component of the digital ballot creation process within the novel framework. This algorithm ensures the integrity and authenticity of the digital ballots and the eligibility of the constituents casting votes. The verification process is depicted in the Business Process Model and Notation (BPMN) and sequence diagrams provided in the related figures.

The algorithm commences with the Election-tech Officer's initiation of the consortium's distributed Election Management System (EMS). The digital ballot creation process involves several critical steps, including the finalization of a quorum security tokens' Hold Your Own Key (HYOK) ceremony and the enabling of AI-based continuous authentication Decentralized Identifiers (DIDs). Once these preliminary steps are completed, a Digi-Ballot claim is issued as a Verifiable Credential (VC) with DID:CCF/Web, which is then multi-signed by the consortium's officers to establish irrevocability.

The eligible constituent, equipped with their Self-Sovereign Identity (SSI) Wallet, verifies the VC. Upon successful verification, the constituent may cast a confidential-tally attempted-vote. The algorithm accounts for coercion-resistance, allowing constituents to produce either a spoiled-vote or a committed-vote. A committed-vote triggers the transfer of confidential-tally and Committed-Vote Zero-Knowledge Proofs (ZKPs) to the final Digi-Ballotbox Audit, ensuring a certifiable audit trail.

The verification algorithm concludes with the transfer of the Digi-Ballot hash to the Zero-Trust Registry-Distributed Ledger Technology (DLT), which includes the eligible-constituent public key address. The final step is the publication of certifiable results and foresight, marking the completion of a secure and verifiable voting process.

The sequence diagrams further elaborate on the interactions between the constituent VC-holder, the SSI-as-a-Service Edge Custodianship Wallet Platform as a Service (PaaS), the Verification-as-a-Service Authenticator, and the Zero-Trust Registry Service Trustee DLT. These interactions are governed by our Critical Infrastructure as Code (IaC) for Trust Over IP (ToIP) framework, ensuring a certifiable, privacy-aware, and scalable i-Voting system.

\vspace{-0.25cm}
\subsection{Constituent Digital Wallet PoC}
\vspace{-0.25cm}
The Constituent Digital Wallet Proof of Concept (PoC) is a critical component of the framework, designed to demonstrate the practical implementation of a secure and privacy-preserving digital wallet system for constituents. This system is integral to the digital ballot creation process and the overall election management system, as it provides a means for constituents to interact with the election infrastructure in a manner that is both secure and verifiable.

The PoC encompasses the design and architecture of the digital wallet, which is built upon the principles of Self-Sovereign Identity (SSI) and operates within a Zero-Trust Architecture (ZTA). The digital wallet serves as a secure repository for constituents' credentials and enables them to perform critical election-related actions, such as verifying their identity, receiving and casting digital ballots, and participating in the election process with a high degree of confidence in the system's integrity.

\begin{figure*}
\centering
\includegraphics[width=1.0\textwidth]{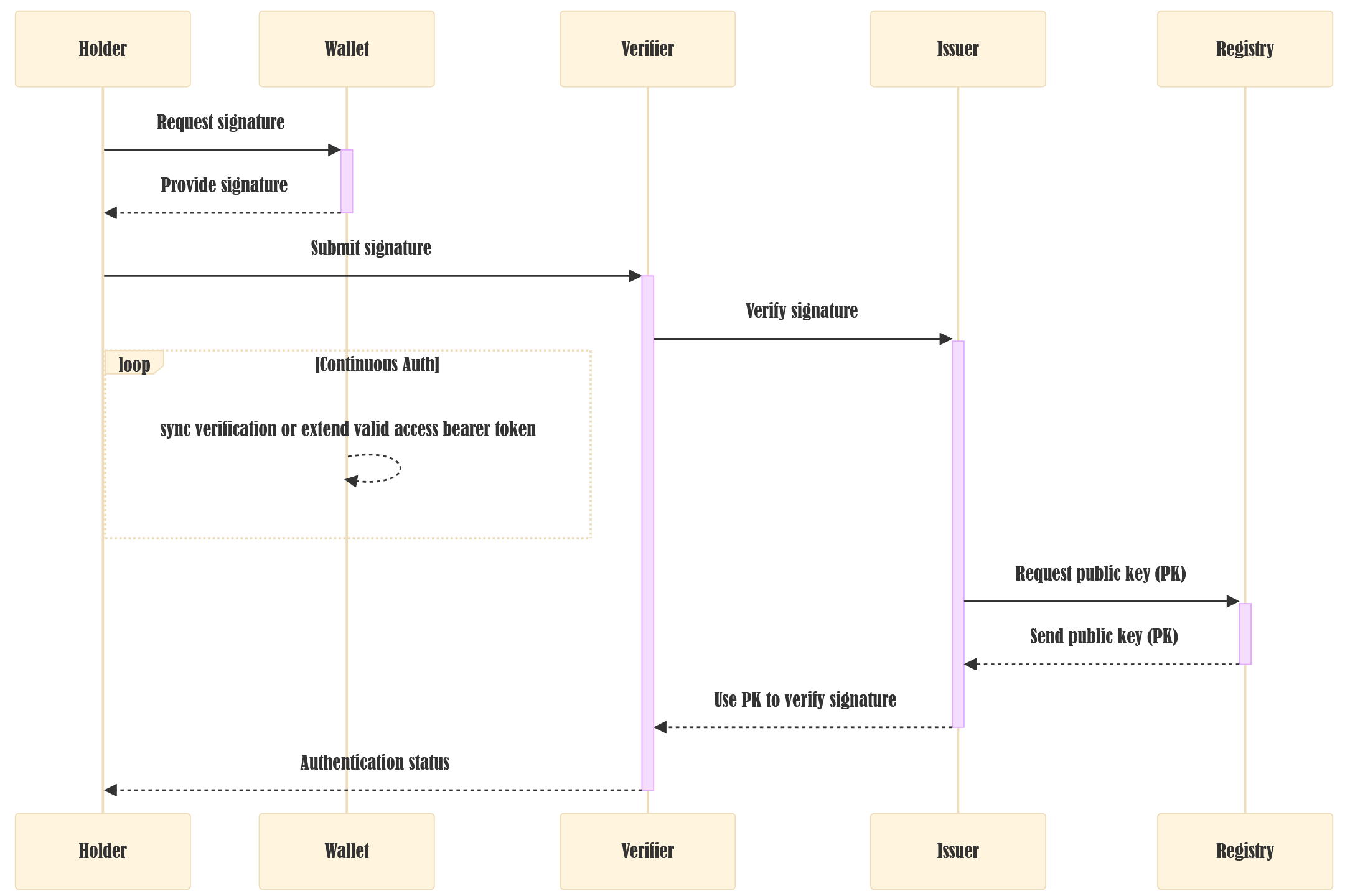}
\vspace{-0.5cm}
\caption{Self-Sovereign Identity (SSI) Sequence Diagram}
\vspace{-0.25cm}
\label{fig:1st-EMS-SSI-ZTA-IaC-ToIP-Seq-Diagram}
\end{figure*}

The sequence diagram in Figure \ref{fig:1st-EMS-SSI-ZTA-IaC-ToIP-Seq-Diagram} outlines the interactions between the constituent VC-holder and various components of the system, including the Self-Attested SSI-as-a-Service Edge Custodianship Wallet Platform as a Service (PaaS), the Verification-as-a-Service Authenticator, and the Zero-Trust Registry Service Trustee Distributed Ledger Technology (DLT). These interactions are facilitated by the Critical Infrastructure as Code (IaC) for Trust Over IP (ToIP) framework, ensuring that the digital wallet operates in a manner that is consistent with the overarching goals of certifiability, privacy, and scalability.

The digital wallet PoC demonstrates the feasibility of the constituent digital wallet system and its ability to integrate with the broader framework. It provides a tangible example of how constituents can securely manage their digital identities and participate in digital elections, thereby contributing to the advancement of certifiable, privacy-aware, and scalable i-Voting systems.

\vspace{-0.25cm}
\vspace{-0.25cm}
\clearpage

\subsection{Business Process Model and Notation (BPMN)}
\vspace{-0.5cm}
\begin{figure*}[!ht]
\vspace{-0.5cm}
\centering
\includegraphics[width=1.0\textwidth]{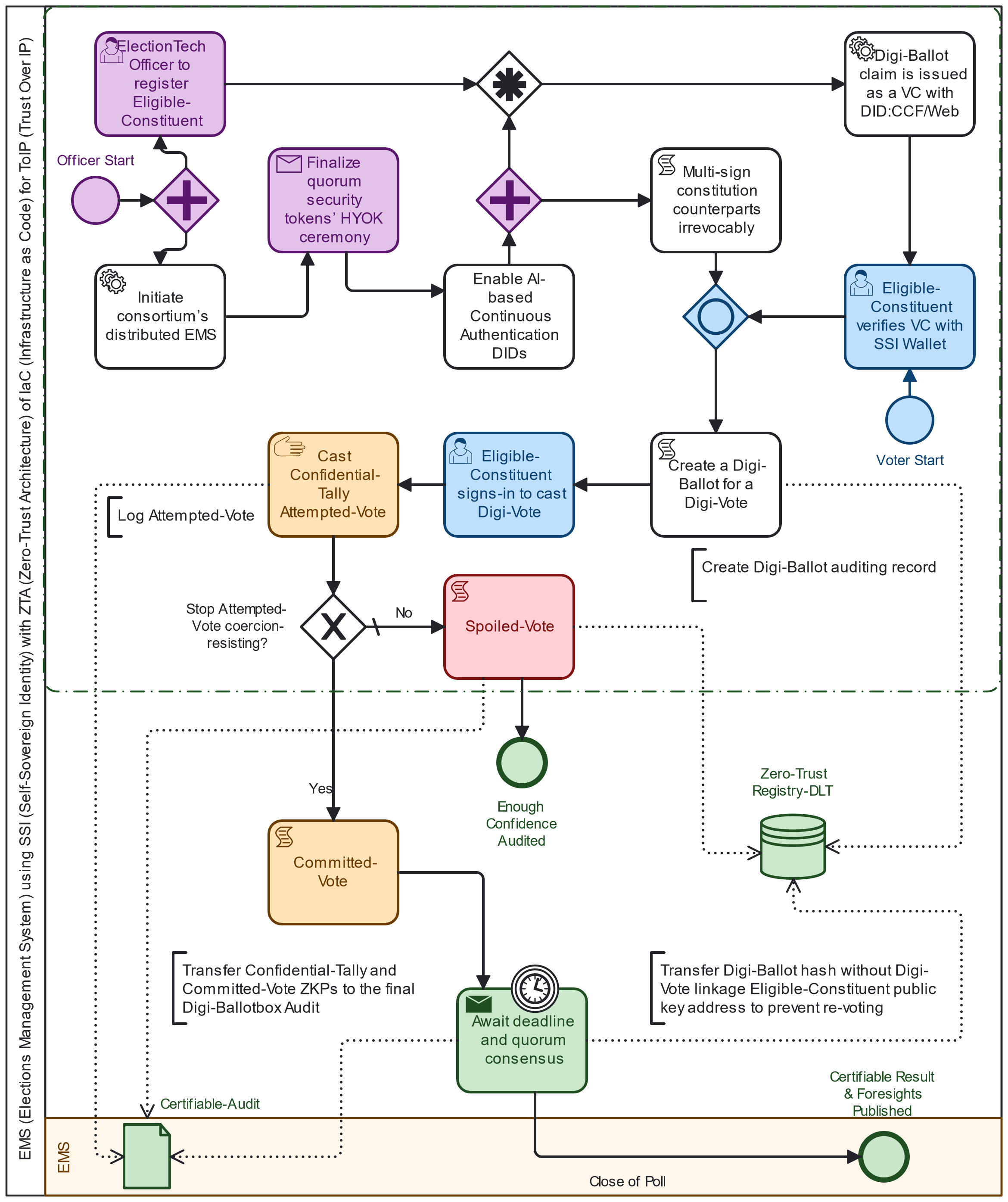}
\vspace{-0.25cm}
\caption{Election-tech Digital Ballot Creation BPMN}
\label{fig:1st-EMS-SSI-ZTA-IaC-ToIP-Electiontech-Digital-Ballot-Creation-BPMN}
\end{figure*}

\vspace{-0.25cm}
\vspace{-0.25cm}
The Business Process Model and Notation (BPMN) provides a graphical representation of the election technology's digital ballot creation and the executed verification algorithm. The BPMN diagram in Figure \ref{fig:1st-EMS-SSI-ZTA-IaC-ToIP-Electiontech-Digital-Ballot-Creation-BPMN} illustrates the sequential flow and decision points that constitute the digital ballot creation process, integrating various components such as the Election-tech Officer, the consortium's distributed EMS, and the Zero-Trust Registry-DLT. The process begins with the Election-tech Officer initiating the distributed EMS and proceeds through several steps, including the finalization of a quorum security tokens' HYOK ceremony, enabling AI-based continuous authentication DIDs, and the issuance of a Digi-Ballot claim as a Verifiable Credential (VC) with DID:CCF/Web.

The BPMN diagram further delineates the path taken by an eligible constituent to verify the VC with their SSI Wallet and cast a confidential-tally attempted-vote. The process accounts for the possibility of vote coercion-resisting, leading to either a spoiled-vote or a committed-vote. The committed-vote then undergoes a transfer of confidential-tally and committed-vote ZKPs to the final Digi-Ballotbox Audit, ensuring a certifiable audit trail. The process concludes with the transfer of the Digi-Ballot privacy-aware hash to the Zero-Trust Registry-DLT, which includes the eligible-constituent public key address to prevent re-voting, and the publication of certifiable results and foresights.

This BPMN is critical for understanding the intricate workflow and ensuring that each step adheres to the principles of certifiability, privacy, and scalability within the context of i-Voting systems. The BPMN serves as a blueprint for the implementation of the digital ballot creation process, ensuring that all actions are verifiable and auditable in a manner that upholds the integrity of the election process within our proposed framework.

\vspace{-0.25cm}
\subsection{App High-Level Design}
\vspace{-0.25cm}
\vspace{-0.25cm}
\begin{figure*}[!ht]
\vspace{-0.5cm}
\centering
\includegraphics[width=1.0\textwidth]{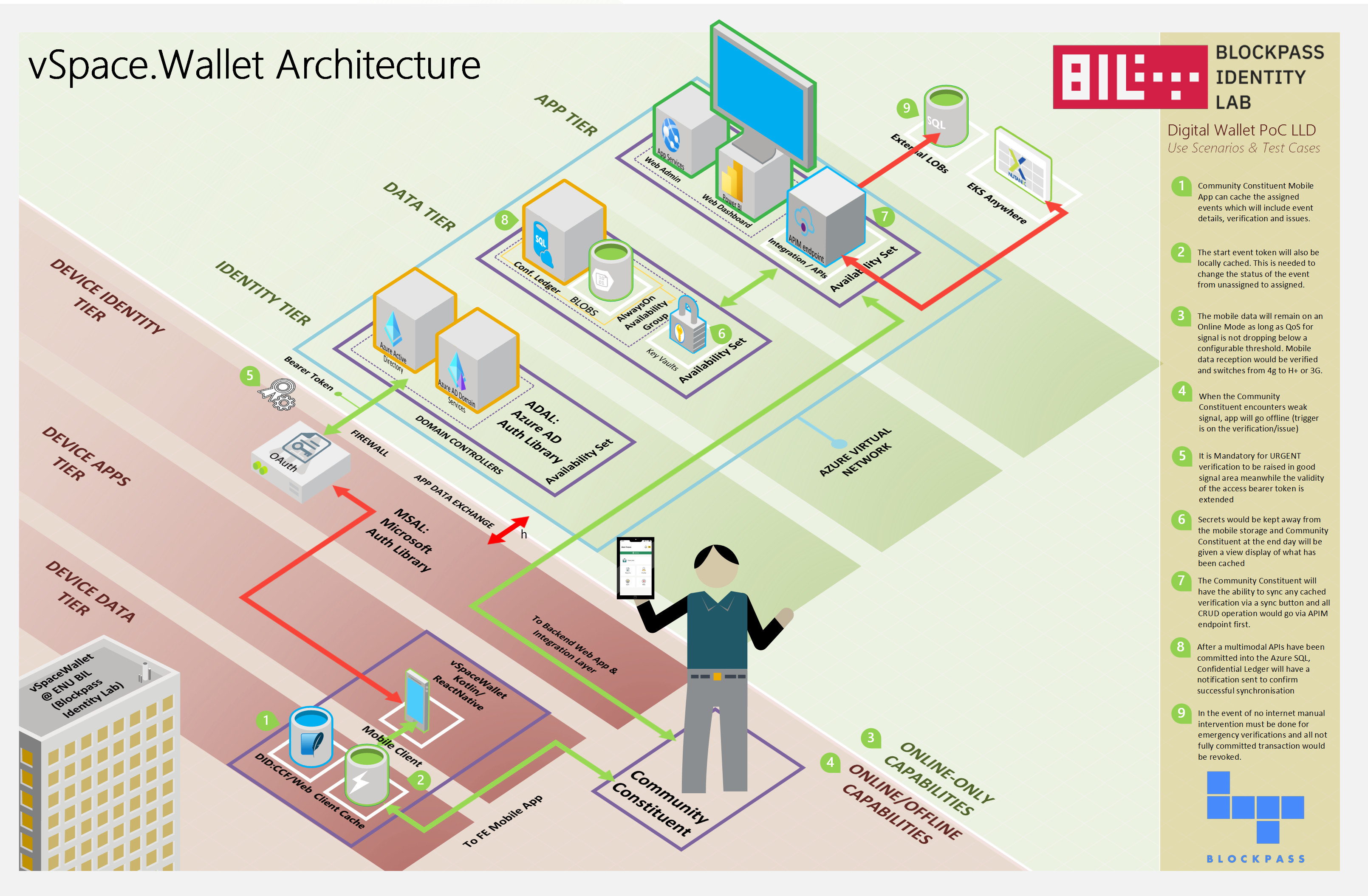}
\vspace{-0.5cm}
\caption{Constituent Digital Wallet App IaC}
\vspace{-0.25cm}
\label{fig:1st-Constituent-Digital-Wallet-ESLZ-App-IaC-LLD}
\end{figure*}
\vspace{-0.25cm}
The envisioned High-Level Design (HLD) is outlined in Figure \ref{fig:1st-Constituent-Digital-Wallet-ESLZ-App-IaC-LLD}. It outlines a scalable digital wallet app implementation leveraging Enterprise-Scale IaC and best practices. By following the AWS's Well-Architected Framework and utilizing modern CI/CD and container orchestration technologies, such Digital Wallet App with a electoral management system (EMS) aims to achieve high availability, security, and scalability for the use case of digital Voting.
\vspace{-0.25cm}




\vspace{-0.1cm}
\section{Evaluation Analysis}
\label{sec:evaluation-analysis}
\vspace{-0.25cm}
ElectAnon is implemented with smart contracts and a ZKP gadget, and it has been observed to run safely for 1,000,000 voters. While the [True{\bf\textit{Elect}}][{\bf\textit{Anon}}Creds] protocol's current evaluation and security analysis are further provided in this section, advanced analysis and more detailed results are pending final insights on our envisioned vSPACE R\&D pilot [Certifiable Consensus using Combined Confidential Computing and Continuous Authentication with Integrity], which aims to achieve certifiable consensus through the integration of confidential computing and continuous authentication with integrity attestation. This system is designed as a SERVE (Secure Electronic Registration and Voting Election-tech) inspired Infrastructure as Code (IaC). This approach is critical for ensuring the security, privacy, and integrity of the voting process, addressing the challenges of insider threats, network vulnerabilities, and auditing difficulties that have been identified in previous studies \cite{gibson2016review} \cite{Kohno2004}.

In addition to these foundational elements, our research is deeply invested in exploring advanced cryptographic techniques, notably Zero-Knowledge Proofs (ZKPs). ZKPs enable a party to prove the truth of a statement without revealing any information beyond the validity of the statement itself. This cryptographic method is essential for a voting infrastructure that prioritizes privacy, verifiability, and transparency while eliminating associated risks. The seminal works of Fiat \& Shamir (1986) and Goldreich \& Oren (1994) laid the groundwork for ZKPs, highlighting their potential in secure cryptographic protocols \cite{feige_zero-knowledge_1988}. 

Our ambition is to pioneer the first technology of its kind that could become a future standard for secure, private, and accessible decision-making processes. This technology is envisioned to be universally applicable, from public engagements and elections to boardroom decisions. By offering an alternative to traditional by-mail systems and current election management solutions, our system aims to enhance the accessibility of i-Voting and/or REV for absentee and overseas citizens \cite{gibson2016review} \cite{Kohno2004}. 

The integration of Distributed Ledger Technology (DLT) for immutable and certifiable audit trails further strengthens the system's integrity and trustworthiness. DLT's application in securing IoT deployments and enhancing the overall security posture of such systems is well-documented, underscoring its potential in our voting infrastructure \cite{Kohno2004}. Moreover, the use of Kubernetes confidential clusters within an Enterprise-Scale Landing Zone (ESLZ) aligns with our commitment to adopting a Zero-Trust Architecture (ZTA) for cybersecurity. This approach ensures that all components of the system are continuously authenticated and verified, minimizing the risk of unauthorized access and data breaches \cite{gibson2016review} \cite{Kohno2004}.


\vspace{-0.25cm}
\subsection{Security Analysis}
\vspace{-0.25cm}
The security analysis of the vSPACE PoC on the [True{\bf\textit{Elect}}][{\bf\textit{Anon}}Creds] protocol is critical in ensuring the integrity, confidentiality, and scalability of digital elections. This analysis leverages the integration of SSI (Self-Sovereign Identity) within a Zero-Trust Architecture (ZTA), employing confidential computing, continuous authentication, multi-party computation (MPC), and well-architected framework (WAF) principles.

The vSPACE framework/PoC employs a Kubernetes confidential cluster within an Enterprise-Scale Landing Zone (ESLZ), integrating Distributed Ledger Technology (DLT) for immutable and certifiable audit trails. The Infrastructure as Code (IaC) model ensures rapid deployment, consistent management, and adherence to security standards, making vSPACE a future-proof solution for digital voting systems

The security assurance focuses on several key areas:
\begin{itemize}
\item \textbf{Certifiable Confidential Computing}: Ensuring that data is encrypted at rest, in transit, and in use. The use of confidential computing technologies, such as CNCF-certified Constellation Kubernetes with aTLS (attested TLS) protocol, provides end-to-end certifiable confidentiality
\item \textbf{Privacy-aware Continuous Authentication with SSI}: Leveraging continuous authentication mechanisms, such as Multimodal Fusion-based Continuous Authentication (MFCA), to ensure that only authorized individuals can access the voting system. This is further enhanced by the integration of Keycloak for SSI OpenID Connect-based authentication
\item \textbf{Secure Multi-Party Computation (MPC)}: Utilizing MPC to enable secure computation over encrypted data, ensuring the privacy of voters' choices while allowing for the accurate tallying of votes
\item \textbf{Distributed Ledger Technology (DLT) Auditability}: Integrating DLT for creating immutable and certifiable audit trails, enhancing the transparency and verifiability of the election process
\item \textbf{Well-Architected Framework (WAF) principles}: Adhering to WAF principles for operational excellence, security, reliability, performance efficiency, cost optimization, and sustainability. This includes passing the CIS Kubernetes security benchmarks and defending the system against various cyberattacks
\item \textbf{ZTA Infrastructure as Code (IaC) compliance}: Employing IaC for rapid deployment and consistent management of the voting infrastructure, ensuring compliance with security standards
\end{itemize}



\vspace{-0.25cm}
\subsection{Protocol Evaluation - Comparative Analysis}
\vspace{-0.25cm}

Table \ref{table:comparison} outlines a comparative analysis of [True{\bf\textit{Elect}}][{\bf\textit{Anon}}Creds] in contrast to related protocols [McCorry et al. (2017), Chaintegrity (2019), Yang et al. (2020), Panja et al. (2020), Priscore, and ElectAnon as evaluated by Onur \& Yurdakul (2022)]:
\begin{itemize}
    \item \textbf{Scalability}: Efficiently managing a large number of voters and candidates without compromising performance or security, enabling scalability according to election needs.
    \item \textbf{Certifiable Confidentiality}: Providing immutable and certifiable audit trails through the integration of off-chain confidential computing and/or for the confidential cluster of an on-chain DLT implementation.
    \item \textbf{DIDs/SSI Continuous Authentication}: Utilizing Self-Sovereign Identity (SSI) for Multimodal Fusion-based Continuous Authentication (MFCA).
    \item \textbf{Consensus AI Hybridization}: Leveraging AI techniques (such as: feature extraction and anomaly detection) for the likes of privacy-preserving biometric authentication, enhanced consensus mechanisms and/or 51\% attack prevention.
    \item \textbf{Integrity Audit Autonomy}: Once initiated, the system should function independently without the need for manual intervention, ensuring an uninterrupted election process; while also ensuring the integrity of the voting process through comprehensive auditability.
\end{itemize}
\vspace{-0.5cm}
\begin{table*}
\caption{Evaluation summary of related DLT-Voting studies.}
\centering
\resizebox{\columnwidth}{!}{%
\begin{tabular}{|c|c|c|c|c|c|c|c|c|c|c|c|c|c|c|c|}
\hline
 DLT-Voting Work &
  \rotatebox{90}{Eligibility Requirement} &
  \rotatebox{90}{Uniqueness Requirement} &
  \rotatebox{90}{Privacy Requirement} &
  \rotatebox{90}{Universal Anonymity Req.} &
  \rotatebox{90}{Fairness Requirement Req.} &
  \rotatebox{90}{Accuracy Requirement} &
  \rotatebox{90}{Universal-Verifiability Req.} &
  \rotatebox{90}{Individual-Verifiability Req.} &
  \rotatebox{90}{Robustness Requirement} &
  \rotatebox{90}{Self-Tallying} &
  \rotatebox{90}{Scalability} &
  \rotatebox{90}{Certifiable Confidentiality} &
  \rotatebox{90}{DIDs/SSI Continuous Auth.} &
  \rotatebox{90}{Consensus AI Hybridization} &
  \rotatebox{90}{Integrity Audit Autonomy} \\ \hline
McCorry et al.\cite{ovnet} &
  \checkmark &
  x &
  \checkmark &
  x          &
  \checkmark &
  \checkmark &
  \checkmark &
  \checkmark &
  o &
  \checkmark &
  x &
  x &
  x &
  x &
  x \\ \hline
Chaintegrity \cite{Zhang2019ChaintegrityBL} &
  \checkmark &
  \checkmark &
  \checkmark &
  x &
  x          &
  \checkmark &
  \checkmark &
  \checkmark &
  x &
  x &
  x &
  x &
  x &
  x &
  x \\ \hline
Yang et al. \cite{YANG2020859} &
  \checkmark &
  \checkmark &
  \checkmark &
  x          &
  x  &
  \checkmark &
  o &
  \checkmark &
  x &
  o &
  x &
  x &
  x &
  x &
  x \\ \hline
Panja et al.\cite{panja} &
  \checkmark &
  x &
  \checkmark &
  x          &
  \checkmark &
  \checkmark &
  \checkmark &
  \checkmark &
  o &
  \checkmark &
  x &
  x &
  x &
  x &
  x \\ \hline
Priscore \cite{priscore} &
  \checkmark &
  \checkmark &
  o &
  x          &
  \checkmark &
  \checkmark &
  \checkmark &
  \checkmark &
  o &
  o &
  x &
  x &
  x &
  x &
  x \\ \hline
ElectAnon \cite{onur2022electanon} &
  \checkmark &
  \checkmark &
  \checkmark &
  \checkmark &
  \checkmark &
  \checkmark &
  \checkmark &
  \checkmark &
  \checkmark &
  \checkmark &
  \checkmark &
  x &
  x &
  x &
  x \\ \hline
[True{\bf\textit{Elect}}][{\bf\textit{Anon}}Creds] (this work) &
  \checkmark &
  \checkmark &
  \checkmark &
  \checkmark &
  \checkmark &
  \checkmark &
  \checkmark &
  \checkmark &
  \checkmark &
  \checkmark &
  \checkmark &
  \checkmark &
  \checkmark &
  \checkmark &
  \checkmark \\ \hline
\end{tabular}%
}
\resizebox{\columnwidth}{!}{%
\begin{tabular}{c} 
\tiny \checkmark: implemented, x: not implemented, o: partially implemented
\end{tabular}
}
\label{table:comparison}
\end{table*}
\vspace{-0.25cm}

\vspace{-0.25cm}
\section{Conclusions}
\vspace{-0.25cm}
This paper has outlined the vSPACE experimental proof-of-concept (PoC) on the [True{\bf\textit{Elect}}][{\bf\textit{Anon}}Creds] protocol. This approach extends the TrueElect and ElectAnon protocols with the integration of Self-Sovereign Identity (SSI). The core advantages include scalability, integrated privacy and confidential computing; not to mention the next future direction of further innovating the credentialing security through Quantum metamaterials and/or AI-based privacy-preserving biometric Verified Credentials (VCs) with Multimodal Fusion-based Continuous Authentication (MFCA).


\vspace{-0.25cm}
\bibliographystyle{IEEEtran}
\bibliography{ref}









\end{document}